\documentstyle[psfig]{mn2e}
\def\ltsima{$\; \buildrel < \over \sim \;$}
\def\simlt{\lower.5ex\hbox{\ltsima}}   
\def\gtsima{$\; \buildrel > \over \sim \;$}
\def\simgt{\lower.5ex\hbox{\gtsima}}

\input{epsf}

\title{The baryonic mass-velocity relation; clues to feedback processes 
during structure formation and the cosmic baryon inventory}

\author{Lucio Mayer $^1$ \& Ben Moore $^1$ \\Institute of Theoretical Physics, University of Z\"urich,  Wintherurestrasse 190, 8057 Zurich, Switzerland}

\begin{document}

\maketitle

\begin{abstract}

We show that a global relation between baryonic mass and virial velocity
can be constructed from the scale of dwarf galaxies up to that of rich galaxy clusters. 
The slope of this relation is close to that expected if dark matter
halos form in the standard hierarchical cosmogony and capture a
universal baryon fraction, once the details of halo structure and the
adiabatic contraction of halos due to cooling gas are taken into account. 
The scatter and deficiency of baryons within
low mass halos ($V_{vir} < 50$ km/s) is consistent with the expected
suppression of gas accretion by photo-evaporation due to the 
cosmic UV background at high redshift. The data are not consistent  
with significant gas removal from strong supernovae winds unless
the velocities of galaxies measured from their gas kinematics 
are significantly lower than the true halo velocities for objects
with $V_{vir} < 100$ km/s. Thus models such as $\Lambda$CDM with a steep mass function 
of halos may find it difficult to reproduce both the baryonic mass-velocity
relation presented here whilst at the same time reproducing the flat 
luminosity/HI function of galaxies. 
Galaxies hold about 10\% of the baryons in the Universe, which is close
to the collapsed mass fraction expected within hierarchical models on these scales, suggesting a 
high efficiency for
galaxy formation. Most of the baryons are expected to be evenly distributed 
between diffuse intergalactic gas in low density environments and the  
intra-galactic medium within galaxy groups.

\end{abstract}

\begin{keywords}{cosmology:theory --- galaxies: dynamics --- galaxies: 
halos}

\end{keywords}

\section{Introduction}

Rotational velocities and luminosities of disk galaxies combine to yield 
the well known Tully-Fisher relation (Tully \& Fisher 1978)
 across several decades of galaxy 
masses. The break in the Tully-Fisher relation at velocities lower than $\sim 90$
km/s is removed once total baryonic 
masses, including gas masses, are used instead of luminosities
(McGaugh et al. 2000). Many faint disk galaxies are indeed gas-rich, with
the neutral hydrogen component often outweighting the stellar mass 
(Schombert et al. 2001). The latter "baryonic" Tully-Fisher is well 
defined down to velocities as low as $50$ km/s, with the small 
intrinsic scatter possibly due to the spread in 
stellar-mass-to-light ratios resulting from reasonable variations in
the star formation histories (Verheijen et al. 1997). The slope of the
baryonic Tully-Fisher measured by McGaugh et al. (2000, hereafter MC00)
is close to $4$. As the relation links the amount of baryons within 
galaxies with their overall potential/total mass (through the rotational 
velocity), it reflects a tight coupling between dark matter and baryons
and hence provides an important test for galaxy formation models. 
Taken at face value the observed slope might be too steep compared to the  slope
of the relation between virial mass and peak velocity of halos expected
in a concordance $\Lambda$CDM model ($\sim 3.5$ - see Bullock 
et al. 2001). However the baryons themselves can modify the mass 
profile as a result of adiabatic contraction (Blumenthal et al. 1986), which 
raises the peak velocity - an effect that must be taken into account when 
comparing theory with observations.

The apparent break existing in the Tully Fisher at low velocities has often been interpreted as evidence for a strong effect of supernovae feedback that
ejects baryons from galaxies (Dekel \& Silk 1986).
The absence of a break in the observed 
baryonic Tully-Fisher relation does not support these feedback models (MC00).
However it is not clear whether this is true for even for fainter
more extreme dwarf galaxies, like those that populate the outer fringes of the Local Group (with measured rotational velocities lower than 50 km/s, see
Mateo 1998). 
Indeed, whilst the most sophisticated numerical models of 
supernovae explosions suggest that even at such low galactic masses only
a very  small fraction of the total gas mass can be removed by supernovae winds
(Mac Low \& Ferrara 1999; Mori, Ferrara \& Madau 2002), the need
to suppress the overcooling in galaxies (White \& Frenk 1991) and
the failure of cosmological simulations with hydrodynamics to form realistic 
disks is 
usually taken as a strong motivation for the need of strong supernovae winds 
(Navarro \& White 1993; Navarro \& Steinmetz 2000; Thacker \& Couchman 2001; but see Governato et  al. 2003). These winds would eject significant amounts of gas in small, early forming objects, quenching galaxy formation at small scales and leaving 
a larger amount of diffuse, higher angular momentum gas available to form
larger galaxies that will be assembled later. A low
efficiency of galaxy formation, suggestive of strong feedback mechanisms,
is also advocated in the recent estimate of the baryonic mass function 
of galaxies (Bell et al. 2003), who find that less than $13\%$ of the 
total number of baryons in the Universe are found in galaxies.

The cosmic UV background at high redshift was strong enough to significantly
suppress the collapse of gas in small halos (Benson et al. 2002a,b, 2003) and
might provide a feedback mechanism capable of explaining why the
number of luminous galaxy satellites of the Milky Way is much lower than
expected from the theory (Kauffmann, White, \& Guiderdoni, 1993; 
Moore et al. 1999; Bullock, Kravtsov \& Weinberg 
2001). On the larger mass scales of groups and clusters of galaxies
strong pre-heating of gas at high-redshift, by
either supernovae or AGNs (Bower et al. 2001; Borgani et al. 2002), 
has also been invoked to 
explain the steepening of the relation between the X-ray luminosity and 
the X-ray temperature of the hot virialized gas
towards decreasing masses (e.g. Babul et 
al. 2000, Borgani et al. 2002). However, there are claims that
radiative cooling alone might account for most of this effect
(Dave et  al. 2001; Bryan 2000). 

In this paper we report on a first attempt to extend the baryonic 
Tully-Fisher relation to both lower masses, by including the faintest
disk galaxies known, and larger masses, up to rich clusters of galaxies.
The implications of our results on the role of feedback mechanisms
in structure formation will be discussed.
We will then revisit the distribution of baryons in the Universe within 
the concordance cosmological model.

\section{Construction of the sample}

We use a variety of datasets to construct the extended baryonic Tully Fisher relation.
(We assume $H_0 = 70$ km s$^{-1}$Mpc$^{-1}$ throughout.)
The total baryonic mass
of systems are inferred directly by observations employing a variety of tracers, 
from HI and mostly near infrared photometry in galaxies to hot ionized gas 
measured through its X-ray emission in clusters.
For bright galaxies our analysis is mostly based on the data published by MC00 (these
are already a combination of different samples, with photometry 
in B, H, I and K bands) to which we add recent B band photometry and HI
kinematics of dwarf galaxies from Stil et al. (2003a,b) and the data on the
outer Local Group 
dwarf irregular galaxies from Mateo (1998). Note that that we do not include the very
nearby dwarf
spheroidals because the structure of these galaxies may have been substantially
reshaped by the tidal interactions with the Milky Way and M31 
(Mayer et al. 2001a,b). We also stress that the baryonic mass estimated for galaxies is more precisely the sum of the stars and the cold gas component
(the latter is the mass of HI augmented by the mass in helium and metals
but no molecular hydrogen, computed as in MC00).  We do
not take into account the eventual contribution of a warm/hot ionized medium
in their halos or disks since the quantitative information from observations is
still poor; however, we will discuss the impact that this would have on the
estimates of the total baryonic content of galaxies in light of recent 
observations in Section 4.

For clusters 
we use two datasets, one from Ettori \& Fabian (1999), which
contains 36 rich clusters observed with ROSAT, and one from Ettori et al. (2002) 
containing 50 clusters with a slightly lower average temperature 
observed by BeppoSax. The same method was used in these two latter papers
to derive cluster masses using fits to NFW profiles. 
For groups we
use the small sample by Mulchaey et al. (1996), that to our
knowledge, is the only one providing an estimate of both gas masses and 
total stellar masses which are non negligible in groups. 
We assume a fixed stellar mass-to-light ratio (in any
given band) to compute the stellar mass from the luminosity of galaxies; we 
follow MC00 (from which the largest sample is drawn), therefore
$(M/L_{*K})=0.8$ and $M/L_B = 1.4$ (these stellar mass-to-light ratios
are based on a stellar population synthesis model orginally developed by de Jong (1996) assuming a Salpeter stellar initial mass function (IMF), see MC00 for details).

We have to make some assumptions to derive the virial circular velocity, $V_{vir}$,  
from  kinematics of galaxies or from the measured temperature of the intracluster medium. 
These assumptions are based on the current paradigm of structure
formation within a cold dark matter scenario. Hereafter we will assume the
standard $\Lambda$CDM model ($\Omega_0 = 0.3$, $\Lambda_0=0.7, \sigma_8=0.9$).
Circular velocity profiles of CDM halos are not flat; they reach a
peak value, $V_{peak}$ at some inner radius and then fall gently 
out to the virial value,
$V_{vir}$. The ratio $V_{peak}/V_{vir}$ depends on the concentration, $c=R_{vir}/
r_s$, where $R_{vir}$ is the halo virial radius and $r_s$ is the halo scale
radius. We have $V_{peak}={[cf(c)]}^{1/2} V_{vir}$, with $f(c)=$4.62[log$(1+c) 
- c{(1 + c)]}^{-1}$ (Bullock et al. 2001).
Kinematical data are normally limited to the inner portion of the galaxies,
hence only $V_{peak}$ is accessible (we discuss later the possibility that
even $V_{peak}$ has not really been measured for many dwarf galaxies).

For galaxies with resolved rotation curves $V_{peak}$ is typically identified with the flat portion 
of the rotation curve, otherwise the half-line width is taken as a reference value (see MC00 and
Gonzalez et al. 2000). $V_{vir}$ can be than computed from $V_{peak}$ by means of the function $f(c)$. 
Cosmological simulations (Bullock et al. 2001)
show that the mean value of $f(c)$ changes by less 
than 30\% between $10^{11}$ and a few times $10^{12} M_{\odot}$ due to the 
mild trend of increasing concentration with decreasing halo virial mass -
the mean value of $c$ varies between $10$ and $18$ in this mass range; 
galaxies with $V_{peak} > 90$ km/s are expected
to have a virial mass larger than $10^{11} M_{\odot}$ (Bullock et al.
2001), and hence for them we assume  $c=14$ as a representative
value to calculate $f(c)$.
The rotation curves of many dwarf and low surface brightness galaxies often
suggest the presence of a constant density core instead of the inner cusp 
of the NFW profile (de Blok, McGaugh \& Rubin, 2001a,b; 
de Blok \& Bosma 2002). However,
here we are not interested in the mass distribution near center of
galaxies, instead we want to estimate the global parameters of a given
system, and in this respect we rely on the fact
that reasonable NFW fits to most of the extent of the rotation curve 
can be obtained provided that one uses concentrations in the range
$3-8$, significantly lower than 
expected in $\Lambda$CDM models at the scale of dwarf galaxies (Van den
Bosch \& Swaters 2001; Swaters et al. 2003a;
Blais-Ouellette, Amram \& Carignan et al. 1999, 2001). 
Therefore, we compute $f(c)$ for a 
fixed $c=5$  for galaxies with $V_{peak} < 90$ km/s (note that typical
concentrations for such systems, whose total mass is supposedly lower 
than $10^{11}M_{\odot}$, should be $\ge 18$ - see Bullock et al. 2001).

We further correct $V_{peak}$ for the steepening of the rotation curve which
would result by the infall of baryons and the adiabatic contraction of the
halo during galaxy formation (Blumenthal et al. 1994). We adopt the
fitting functions by Mo, Mao \& White (1998), which  depend on the halo spin
parameter $\lambda$, $c$, the disk mass fraction $f_d$ and the ratio between
disk and halo specific angular momentum, $j_d/j_h$. We assume $j_d/j_h = 1$,
namely that dark matter and baryons start with the same specific angular momentum and baryons conserve the latter during collapse.  
Assuming the most probable value for the halo spin 
($\lambda=0.035$, see Gardner 2001) and a conservative
value for the disk mass fraction $f_d=0.05$ 
(e.g. Jimenez, Verde \& Oh 2003), this last correction lowers by another 
$20\%$ the value of $V_{vir}$ calculated from $V_{peak}$. 
For simplicity 
we assume a single correction factor for all galaxies with $V_{peak} 
> 90$ km/s (corresponding to $V_{vir} \sim 50$ km/s). 
We do not apply the correction for adiabatic contraction  
in  galaxies with lower $V_{peak}$ - indeed photoionization at 
high redshift should have reduced substantially the infall of baryons
within small halos, leaving their dark matter circular velocity profiles
nearly unaffected by the baryons (Quinn, Katz \& Efstathiou 1996; Gnedin 2000).


Finally, when its contribution to the kinematics is non-negligible (typically
for the faintest dwarf irregulars) we also include the gas velocity dispersion
in the calculation, defining $V_{peak}= \sqrt{{V_{rot}}^2 + 
\beta  \sigma^{2}}$ ($\sigma$ is the 1D, line-of-sight velocity dispersion,
$V_{rot}$ is the rotational velocity), which follows from the virial theorem 
(Swaters et al. 2003b), and we assume isotropy such that $\beta=3$.

For clusters and groups we use X-ray temperatures of the diffuse  hot 
gaseous medium to infer the 1-dimensional
velocity dispersion, $\sigma$, under the assumption that the system is 
in virial equilibrium, $T_{vir} \simeq 0.13 {\sigma}^2 \mu m_p/k_B$
(see Binney \& Tremaine 1987), where  
the molecular weight is $\mu=0.5989$ (we assume ionized gas with 
cosmological abundances) and $m_p$ is the mass of the proton.
The velocity dispersion is then used to 
determine the circular velocity by simply assuming the asymptotic relation 
valid for an isothermal potential, $V_{vir} \sim \sqrt{2} \sigma$, which is
approximately valid even for an NFW profile (Taffoni et al. 2003).
We use the gas masses measured within the outermost radius
for all clusters; this radius is between 1 and 1.5 Mpc and we assume
that it is a good estimate of the virial radius (if the true virial radius
is larger we should only slightly underestimate the total gas mass given the
steep outer slope of the NFW profile). 

For some of the groups and clusters 
it is possible to compare the masses inferred 
from using the X-ray data and optical velocity dispersion data. We found 
that the agreement is very good for all clusters while for some
groups, especially those whose X-ray emission is not centered and smooth,
the resulting dispersions are smaller than those derived from
kinematics, which in turn results in smaller virial masses. When the
disagreement is strong we remove the group from the sample as this
might indicate an unbound or, at least, non virialized system.
We caution that the groups are the most uncertain among the datasets; 
the extent of the X-ray emission is limited by instrumental sensitivity and 
in general a smaller fraction of the virial 
radius is probed (Mulchaey \& Zabludoff 1998).
As a consequence, the estimated gas masses for groups are simply 
lower limits.

\section{The extended baryonic mass-velocity relation}

In Figure 1 we show that a baryonic mass-velocity relation  holds
across the entire range of scales of virialized objects. The line shown
follows the expected mass-velocity relation of dark halos in a $\Lambda$CDM model, where $M_{vir} \sim {V_{vir}}^{3}$. To derive the latter we
calculate the baryonic mass $M_{bar}$ at any given value of the circular 
velocity $V_{vir}$ as $M_{bar}=f_b M_{vir}$, where $f_b$ is the universal 
baryon fraction, whose best estimate is $f_b = 0.17$ (Spergel et al. 2003), 
and $M_{vir}$ is the virial mass at a given $V_{vir}$ expected for virialized halos
in a standard $\Lambda$CDM model.

As shown in Figure 1, data and theory can be brought into a reasonable 
agreement once the correction for both the added baryonic mass and the 
adiabatic contraction of the halo are properly taken into account, 
contrary to previous claims (MC00).
We stress that applying the correction for the adiabatic contraction of the 
halos is essential to reach consistency with the theoretical curve at galaxy 
scales. We also note that the correction accounting
for different halo concentrations depends on
the normalization of the power spectrum, hence on
 $\sigma_8$.  Here we assumed  $\sigma_8=0.9$, lower/higher values
will yield less/higher concentrated halos and thus a smaller/bigger
correction to the observed $V_{peak}$, respectively.
Although a mean relation exists, 
the data deviate from the simplest theoretical prediction at group scales (near
$V_{vir} = 300$ km/s) and at the scale of the
smallest dwarf galaxies, corresponding to $V_{vir} < 50$ km/s (in 
particular, the best-fit curve at small scales would have a steeper slope, 
around -3.4). 
In both 
cases the deviation
can be seen as a deficit of baryons at a given value of the circular velocity (the
opposite interpretation, namely an overestimate of the circular velocity, is highly unlikely,
at least for galaxies, as the observed velocities have been reduced as much as
possible following the assumption that the data
yield $V_{peak}$ - if some of the rotation curves are still rising we would 
be underestimating $V_{vir}$).

The deviation and increased scatter 
at dwarf galaxy scales can be easily explained as a result
photoionization by the UV background at high redshift. 
Semi-analytical models and
numerical simulations  (Quinn, Katz \& Efstathiou 1996; Benson et al. 2002a,b; 
Bullock et al. 1999; Thoul \& Weinberg 1996) suggest that
gas collapse might have been substantially inhibited for objects with 
$V_{vir} < 50$ km/s once reionization begins. 
At even lower circular velocities evaporation of gas 
that had already collapsed might also take place (Barkana \& Loeb 1999;
Shaviv \& Dekel 2003).
These previous results may need some re-interpretation in
light of a possible early epoch of reionistation suggested by WMAP 
(Spergel et al. 2003) 

In Figure 1 we compare our results with the predictions from
some of the highest resolution simulations of early galaxy
formation that include the cosmic UV background (Tassis et al. 2003).
We observe a good agreement between the observations and simulations 
in both the scatter and deficiency of baryons within small galaxies.

Simulations from the same authors that also include the effect of
thermal and kinetic heating by supernovae find that
the minimum baryonic masses would be up to three
orders of magnitudes lower than shown in Figure 1. This is also 
similar to the semi-analytic model predictions discussed later.

One could argue that our analysis is missing galaxies with very 
low baryon fractions simply because they would be too faint to be seen. 
These objects might be purely gaseous
or nearly optically dark; a significant population of the gas-rich objects
in the local Universe seems to be ruled out by recent wide-field HI surveys
(Zwaan et al. 2003), but the second possibility cannot be excluded at the moment.

The simulations of Tassis et al. (2003) that include strong
supernovae feedback predict that even fairly bright spiral galaxies,
with masses well in excess of $10^{11} M_{\odot}$, corresponding to
$V_{vir} \ge 100$ km/s, would have an average baryon fraction almost
an order of magnitude lower than the cosmological value, lying well
below the relation reported in Figure 1.

Therefore our results suggest that supernovae winds 
do not eject significant baryonic mass from galaxies.
This, however, does not mean that feedback is not important
as a regulating mechanism for the ambient gas temperature and density, and thus
for star formation, in galaxies both small and large. 
\begin{figure}
\epsfxsize=9truecm
\epsfbox{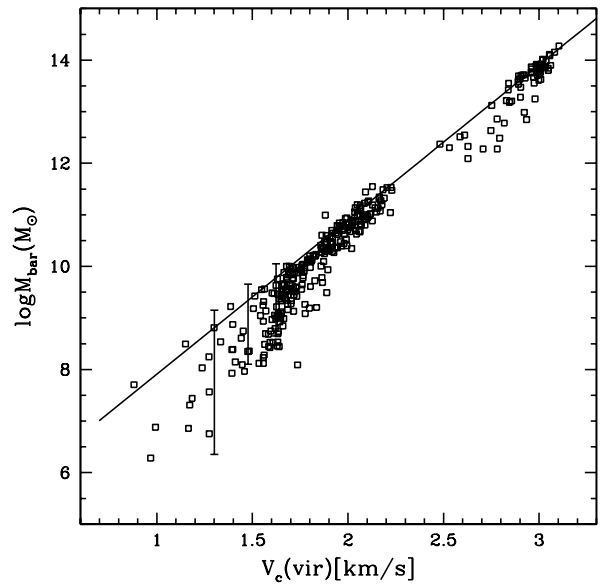}
\caption {The baryonic mass-velocity relation. Data points (squares) have 
been corrected along the velocity axis as described in the text. 
Solid line: theoretical relation between virial mass and virial
velocity predicted by the standard $\Lambda$CDM model (a top-hat collapse model
has been used). 
The error bars show the spread of baryonic masses
at a given halo circular velocity according to the simulations of
Tassis et al. (2003) that include photoionization but no supernovae 
feedback. }
\end{figure}

It is notable that the baryonic Tully-Fisher has such a small scatter across most of the galaxy population. 
As already pointed out by MC00, variations of the stellar mass-to-light ratio due to different
star formation histories would already account for most of the scatter 
along the vertical axis, leaving little room for variation in the IMF of stars.
Along the 
horizontal axis, a scatter of 0.4 in log$(V_{vir})$ would be expected  if, at a fixed value
of the concentration, we vary $\lambda$ in the range $0.01-0.1$ and $f_d$ 
in the range $0.01-0.15$. 
These variations in the main parameters controlling disk formation
inside dark halos already account for the entire scatter in the plot at  
$V_{vir} \sim 100$ km/s.


Cosmic scatter in the structure of dark halos alone, which translates into 
a possible range for the concentration of halos at a given mass, is expected to produce an additional scatter of roughly 0.2-0.3  in log($V_{vir})$ (Bullock et al. 2001), and therefore globally we would expect data points to be more 
scattered than they actually are.
A similar problem was already argued by Bullock et al. (2001) for the Tully-Fisher relation.
However, at least in our datasets, the galaxies considered are only late-type objects. Spheroidal
components are never dominant and this eliminates a large portion of the 
available
parameter space, and thus of the scatter. In particular, both low spin 
objects ($\lambda < 0.03$)
and systems with very high disk mass fractions ($f_d \ge 0.1$) may transform into
early-type spirals or S0 galaxies as a substantial fraction of their disk mass transforms
into a bulge because of bar formation and secular bar evolution 
(Combes et al. 1990; Mo et al. 1998).
Considering the restricted parameter space ($0.03 < \lambda < 0.1$, $0.01 <
f_d < 0.1$), the scatter along the horizontal axis due to variations in
the conditions of disk formation reduces to less than 0.2 in log$(V_{vir})$, 
leaving room for the other possible sources of scatter.

The deviation at group scales is also interesting, although the
interpreation is hindered by the small size of the sample considered
here.  One possibility is that groups contain a substantial mass of
gas at temperatures $10^6--10^7$K that has not been observed because 
it falls below the detection limits of current instruments (Mulchaey
\& Zabludoff 1998;Burstein \& Blumenthal 2002).
%
Alternatively, pre-heating and evaporation of gas induced by winds from AGNs, 
with an  effective reduction of the gas masses bound to the groups, can be 
invoked (Silk \& Rees 1998; Bower et al. 2001).
Indeed, in a scenario where there is a strong link
between the formation of spheroids and supermassive black holes 
(Ferrarese \& Merritt 2000), we
can imagine that X-ray bright groups like those considered here would be 
affected most.
We also note that even at cluster scales several points lie slightly
below the theoretical curve. This might indicate that some
fraction of the baryonic matter is in a warm undetected phase even at these
scales, as recently argued by Ettori (2003).

\section{The baryon pie}

If galaxies have most of their baryons locked in their disks 
it might seem odd that observational measurements of the baryonic mass function
of galaxies indicate that the latter contribute only a tenth
of the total amount of baryons expected in the Universe (e.g. Bell et al. 
2003). However, the question here is how large a contribution do galaxies make
to the total (dark + baryonic) mass of the Universe in the first place?

We use a large high-resolution N-Body simulation to estimate the contribution
of different mass scales to the total mass in a representative volume of the
Universe. The $\Lambda$CDM simulation (Reed et al. 2003) has a box of side 
50 Mpc$^{-1}$ and the particle mass is $1.3 \times 10^{8} M_{\odot}$, such that
it has  enough resolution to probe objects as small as the most massive
dwarf  galaxies in the Local Group (a few times $10^9 M_{\odot}$). 

At $z=0$ we integrate the mass function in different mass bins (Figure 2) and
find for the following broad mass scales:

Galaxies: $10^{10}M_\odot< M_{vir}< 10^{12}M_\odot$ 13\%.

Groups:$10^{12}M_\odot< M_{vir}< 10^{14}M_\odot$ 30\%.

Clusters $M_{vir} > 10^{14}M_\odot$ 10\%.

Note that among the galaxies we have not included bound systems with masses
$M_{vir}  < 10^{10} M_{\odot}$. These are found in the simulation and contribute another $\sim 5 \%$ to the total mass. However, even assuming that they
have a cosmological baryon fraction they would have baryonic masses lower 
than the lower limit in the analysis of Bell et al. (2003).
In addition, as we explained 
above, at these mass scales ($V_c < 40$ km/s) the effect of photoionization 
is important - gas that might have collapsed at these scales 
will more likely end up contributing to a diffuse IGM component (see below).

Bell et al.(2003), by measuring the mass in stars and cold gas within 
galaxies (hence at the baryonic  mass in their disks), find that the 
contribution of  galaxies to the baryon budget is around $8 \pm 5\%$ and interpret this as a  low efficiency for galaxy formation. However, this number is 
quite close to the 13\% that we would estimate here for the expected 
contribution of galaxies to the baryonic pie under the assumption that they captured the 
cosmological baryon fraction. In fact, galaxies will contribute a fraction
$f_{b,gal}= f_x M_{gal}/f_b M_{tot}= (f_x/f_b)f_{M,gal}$ to the
total baryonic content of the Universe, where $f_x$ is the baryonic fraction in galaxies, $f_b$ is the cosmological baryonic fraction, and $f_{M,gal}=M_{gal}/M_{tot}$
is the fractional mass contribution of galaxy-scale objects to the total.
If we assume $f_x=f_b$, using the above estimate for $f_{M,gal}$, namely
$13 \%$, it also follows that  $f_{b,gal}= 13 \%$. This actually
indicates a high efficiency for galaxy formation.

It is likely that galaxies have a substantial component
of hot gas in an extended halo, material that is still cooling inwards onto the disk.
Evidence for the existence of this component is gradually accumulating, 
at least for the Milky Way, thanks to new observations of OVI 
and X-ray absorption (Sembach et al. 2003a,b; 
Nicastro et al. 2003; Kalberla \& Kerp 2001). These observations suggest that
the hot gas could have a density of up to $10^{-4}$ atoms cm$^{-3}$ between 50
and 100 kpc and that its temperature at these distances is less than
$2\times 10^6$ K. Further evidence for a hot halo with this density comes 
from the hydrodynamical model for the LMC-Halo interaction and 
the Magellanic Stream (Mastropietro et al., 2003). 
Assuming that the hot gas profile follows the dark matter (NFW) profile
its total mass would be
about 30\% of the disk mass, or 25\% of the sum
of both components.  If $25\%$ of the galactic baryons are in the this
diffuse halo component, then the fraction of baryons locked into the cold phase
, namely in the disk of stars and gas, will be less than $10\%$, in even betteragreement with the estimate of Bell et al. (2003), who are indeed 
neglecting any baryonic component outside the disks.

\begin{figure}
\epsfxsize=7truecm
\epsfbox{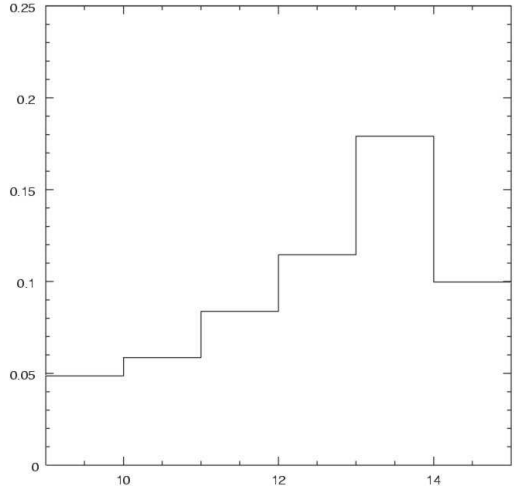}
\caption{Histogram of the mass fraction in objects of different mass scales 
in the high-resolution $\Lambda$CDM simulation of Reed et al. (2003) (see text).}
\end{figure}

Groups of galaxies potentially hold the largest fraction of baryons in
virialised structures, whilst clusters of galaxies (defined
by mass above) would contribute
only about $10\%$. We caution that, due to the modest box size,
statistical $1\sigma$ Poisson fluctuations in the mass function of
objects are of order $30\%$ at the group scale and up to $\sim 80 \%$
at the cluster scale (see Reed et al. 2003). Even so, groups would
always be the most important contribute most of the mass.  The
importance of groups for the baryon budget has been noted by many
authors in the past, among them Fukugita, Hogan \& Peebles (1998).

It is now apparent why clusters and groups contain more baryons in gas
than the sum of the galaxies that formed these systems. The volume from 
which clusters collapse is large enough to capture a large fraction of the 
low density IGM thus giving a final high fraction of diffuse gas.

In the $\Lambda$CDM simulation there 
is also about $42 \%$ of diffuse mass, which
is outside of (resolved) virialized structures.  While this diffuse dark matter
component would certainly collapse in smaller structures at even
higher resolutions (Moore et al. 1999) most of these small halos are
expected to be dark, thereby not contributing to the baryon budget.
Indeed, within halos of masses $10^3 M_{\odot} < M_{vir} < 10^{6} M_{\odot}$
(the lower limit is given by the cosmological Jeans mass) baryons can
cool via molecular hydrogen at very high redshift ($z \ge 25$) but
would immediately photo-dissociate H$2$, halting baryonic collapse
until they achieve masses in the range $M_{vir} > 10^6 M_{\odot}$
(corresponding to a virial temperature $T_{vir} > 10^4$ K) and can cool via atomic hydrogen 
(Haiman, Thoul \& Loeb 1996; Haiman, Abel \& Rees 2000; Haiman 2003).

At later times low mass 
halos may reionize the intergalactic medium, suppressing the
collapse of baryons at scales up to $M_{vir} \sim 10^9 M_{\odot}$. Therefore
the diffuse mass in our simulations should mostly trace a truly
diffuse IGM baryonic component.  This latter component together with gas
inside, or eventually, expelled from groups by AGNs should make up the
dominant contribution to the baryon budget, about $75 \%$ according
to our numbers. A substantial amount of
"warm" gas ($10^5$ K $< T  < 10^6$ K) outside virialized structures would
indeed explain the soft X-ray background (Cen \& Ostriker 
1999; Dave et al. 2001). 
The same reasoning and baryon fractions in the different components would
also apply to warm dark matter, or other models that have reduced power on 
small scales (below $10^{10}M_\odot$).

\section{Discussion}

We have shown that a relation between the mass of baryons and the depth
of the potential well holds across a wide range of scales, from the smallest 
dwarf galaxies to galaxy clusters. The mean relation is consistent with
the mass-velocity relation expected for most cosmological models in which
dark matter halos grow and collapse through gravitational instability.
Deviations from the mean relation at the scale of dwarf galaxies are 
explained as 
a result of heating/evaporation from the UV background at high redshift, while 
at group scales we cannot exclude a role of feedback from AGNs (Silk \& Rees 1998; Kaiser \& Binney 2003).
Our results argue against the existence of the "strong form" of 
supernovae feedback, namely that capable of substantial removal of baryons
in dwarfs (Dekel \& Silk 1986; Dekel \& Woo 2003).

We believe that it will be an interesting challenge for the standard
concordance $\Lambda$CDM model to reproduce both the baryonic mass
function presented here whilst also producing a luminosity (and HI) function of
galaxies with a reasonably flat faint end slope. Most models in which the
dark matter is a collisionless component predict a mass function
of dark matter halos which is much steeper than the luminosity
function of galaxies. These models rely on strong feedback to give rise
to a mass dependent mass-to-light ratio to flatten the observed luminosity
function of halos -- photoionization alone is not enough (see
Benson et al. 2003).

\begin{figure}
\epsfxsize=9truecm
\epsfbox{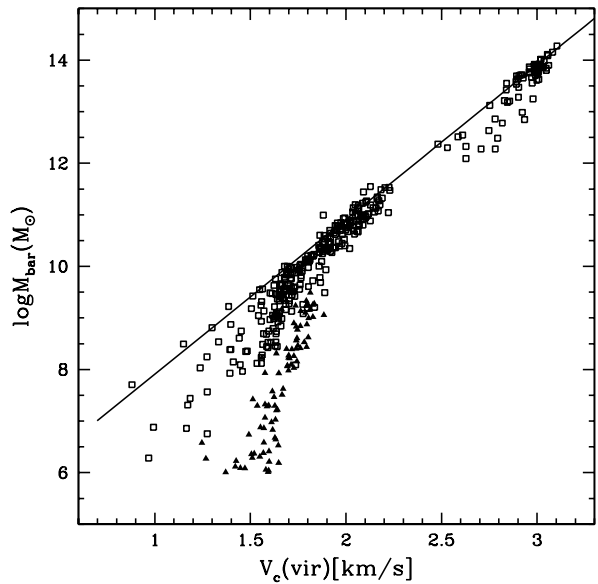}
\caption{Data points (squares) and theoretical relation (solid line) 
compared with the results of the Benson et al. (2002a,b) semi-analytical model 
of galaxy formation  (triangles). The latter model (see Discussion) 
includes photoionization plus a strong kinetic feeback for small halos.}
\end{figure}

In Figure 3 we compare our results with the predictions of the
Durham semi-analytic galaxy formation models that include both
photoionization and supernovae winds (Benson et al. 2002a,b; Cole et
al. 2000). 
In the latter model, a large fraction of the energy of supernovae
explosions is converted into kinetic energy, suppressing gas cooling and
star formation in halos with low values of $V_{peak}$, as in the
numerical simulations of galaxy formation by Navarro \& White (1993).
This has the expected result of reproducing reasonably well the faint end of
the galaxy luminosity function.
This form of feedback is not as strong as the superwinds in Benson et
al. (2003), which can remove gas even in bright ($L_*$) galaxies,
but still produces objects whose baryonic content falls short of that
predicted by the baryonic mass-velocity relation (see Figure 3).
The same models do indeed provide a better fit (within a factor of 2)
to the I-band Tully-Fisher relation, which of course uses only the luminosities
of galaxies (see Figure 7 in Cole et al. 2000); indeed strong feedback will 
remove most of the gas in dwarf galaxies - a larger discrepancy shows up 
in the baryonic Tully-Fisher simply because gas accounts for most of the
baryons in observed dwarfs (MC00). 
We note
that models with a truncated power spectrum at small mass scales such as
would be produced by free streaming of a kev particle or through an
interaction between the dark matter and photons (Boehm et al 2002) might be 
able to reproduce these observed correlations. These models should preserve
the same scaling properties that allowed us to fit the
baryonic mass-velocity relation down to galaxy scales, but would naturally
lower the number of low mass halos such that the mass function has a linear
relation to the luminosity function.

A caveat in the results presented here is that measurements of both the 
peak velocity and the baryonic masses of galaxies are subject to several 
uncertainties, especially in the case of dwarf galaxies. A factor of two 
variation in  the stellar
masses of galaxies is indeed easily achieved by changing the IMF of stars
(Cole et al. 2001). In additions, the data
for the faintest galaxies included in our sample (Stil \& Israel 2002a,b; 
Mateo 1998) do not extend far from the center such that some have rotation curves that 
are not clearly flat at the last measured point. In the smallest galaxies the velocity field of the gas is quite chaotic and is dominated by random motions
in the outer part (for example GR8, see Carignan, Freeman \& Beaulieau 1991) 
such that the association of the measured velocity with 
$V_{peak}$ is uncertain. 
In these cases we cannot exclude that we are only probing the inner part
of a much bigger system with much higher velocity, which would move the
data points to the right in Figure 3, towards the predictions of the 
semi-analytical models. 
A similar argument has been made by Stoher et al. (2002) to fix the
comparison between the observed number of galactic satellites and that
predicted in the $\Lambda$CDM model. As a simple exploration of
where the data points would lie if we push the systematic effects in favour
of cold dark matter models, in  
Figure 4 we show the data points 
after allowing both a factor of 2 increase in the true halo virial velocity
(this being quantitatively consistent with the predictions of Stoher et al.)
and a factor of 2 decrease in the stellar mass of galaxies due to a
different IMF - the correction to the velocity is applied only to galaxies
with measured velocities $<50$ km/s since these have the more poorly
determined rotation curves. In this case there is a much
better agreement with the predictions of semi-anaytical models, but
still not a perfect overlap.

\begin{figure}
\epsfxsize=9truecm
\epsfbox{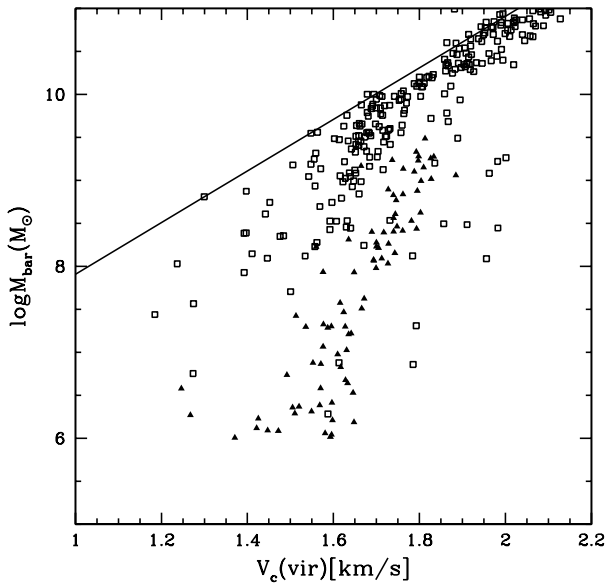}
\caption{As Figure 3, but only showing the comparison for the lower
mass systems  and for virial velocities set equal to twice the measured 
maximum velocity (see Section 4 for discussion on this correction) 
in the case of galaxies with $V_{rot} < 50$ km/s. Stellar masses are also 
reduced by  a factor of 2 to account for possible uncertainties in the
IMF.}
\end{figure}

How does our Galaxy fit in the picture presented so far?
According to the results of Figure 1, the Milky Way, in order to be 
"typical" for a baryonic mass $\simlt 10^{11} M_{\odot}$, as suggested by
its K-band luminosity (Kochanek et al. 2002), must have $V_{vir} \simlt 130$ km/s, 
and a total virial mass $\sim 10^{12} M_{\odot}$. 
Such a model for the Milky Way is plausible based on its rotation curve 
and on the other observational constraints available, 
and in particular, is in agreement with the most baryonic dominated,
maximum disk models (Klypin, Zhao \& Somerville 2002;
Wilkinson \& Evans 1999). The additional $3\times 10^{10}M_\odot$ of 
hot gas in the halo suggested by the LMC kinematics, FUSE and ROSAT data 
would imply that we have accounted for all the expected baryons in the 
Galaxy.

Our analysis does not include individual elliptical galaxies (these of
course enter in the global estimates of baryonic masses in groups and clusters). Interestingly, a recent paper by Padmanabhan et al. (2003), which uses
photometry and kinematics of almost 30.000 elliptical galaxies with velocity
dispersions larger than $70$ km/s taken from the Sloan Digital Sky Survey,
finds that the dynamical to stellar masses are between 7 and
30. Taking into account that a typical elliptical galaxy also has a significant
hot gaseous X-ray halo, they conclude that these galaxies appear to have 
captured close to the cosmological baryon fraction, in agreement to what we
find here for other types of galaxies.

An additional piece of this complex puzzle is the relationship with the 
dynamical mass estimates of field galaxies from weak lensing
(McKay et al 2001, Guzik \& Seljak 2002). 
Weak lensing should provide the strongest
constraints on the total mass-to-light ratios of galaxies. 
McKay et al. (2001) obtained extremely high average mass-to-light ratios,
roughly around $100$. However, more recently Guzik \& Seljak reanalysed
the same SDSS data set taking into account the effects of clustering and 
cosmologically motivated models for the halo density profiles. At $L_*$, 
they find a virialised dark matter halo to baryon mass ratio of 10. They 
also comment that this implies a high
efficiency in the conversion of baryons to stars. In other words, the weak 
lensing data also imply that galaxies have captured the expected baryon 
fraction and 
that feedback has been inefficient at preventing star formation and has not
ejected a large fraction ($>30\%$) of baryons into the IGM.

\bigskip

We thank Darren Reed, Fabio Governato and Tom Quinn for allowing us to use
their $\Lambda$CDM simulation, Andrew Benson, Carlos Frenk and the other members of the Durham theory group for providing us with the results
of the Durham semi-analytical models and for stimulating discussions.
We also thank Frank Van den Bosch and Andi Burkert for helpful comments on an earlier version of the paper and George Lake for numerous discussions on 
feedback.

\end{document}